\documentclass{appolb}

\newcommand{\be}{\begin{equation}}
\newcommand{\ee}{\end{equation}}
\newcommand{\bea}{\begin{eqnarray}}
\newcommand{\eea}{\end{eqnarray}}
\newcommand{\non}{\nonumber}
\newcommand{\Lg}{{\rm ln}}

\newcommand{\ep}{\varepsilon}

\newcommand{\nobody}{\rule{0ex}{1ex}}

\begin{document}
  \eqsec  
\title{{Analytic $\varepsilon-$ Expansion of the \\
Scalar One-loop Bhabha Box Function}
\thanks{Presented at XXVII Conference of Theoretical Physics:
``Matter to the Deepest'', Ustro\'n, Poland, 15-21 September 2003}%
$\nobody^,$%
\thanks{
Work supported in part by European's 5-th Framework
  under contract HPRN-CT-000149.}}

\author{J.~Fleischer
\address{Fakult\"at f\"ur Physik, Universit\"at Bielefeld \\
Universit\"atsstr. 25, D-33615 Bielefeld, Germany }
\and
T.~Riemann, O.V.~Tarasov
\address{Deutsches Elektronen-Synchrotron DESY\\
Platanenallee 6, D--15738 Zeuthen, Germany}
}
\maketitle
\begin{abstract}
{We derive the first three terms of the $\varepsilon-$ expansion of the
scalar one-loop Bhabha box function from a representation in terms of
three generalized hypergeometric functions, which is valid in
arbitrary dimensions.
}
\end{abstract}




\section{{Introduction}}
{
One of the {important} problems in perturbative calculations is a precise
determination of the cross section for Bhabha scattering.
For this one has to determine the electroweak one-loop corrections in the
Standard Model {and some parametric enhanced contributions plus the
complete photonic corrections} to even higher orders.
 {Here} we are interested in a determination of photonic
${\cal O}(\alpha^2)$ corrections for this process in $d=4-2\varepsilon,
{\varepsilon \to 0},$
dimensions with account of the electron mass $m$ as a regulator of
infrared singularities.
These corrections naturally  {concern} the virtual two-loop matrix
element, which
contributes to the cross section due to its interference with the Born matrix
element.
Of the same order is the absolute square of the one-loop amplitude $M_1$.
The corresponding cross section contributions have been analytically
determined recently \cite{Fleischer:2002wa}.
}

{
A peculiarity of the contribution from $M_1$ is the necessity to determine
this amplitude as a function of the parameter $\varepsilon$ up to terms of order
$\varepsilon$: $M_1(\varepsilon)= m_1/\varepsilon + m_0 + m_1 \varepsilon$.
The available numerical programs for the evaluation of one-loop functions are
not sufficient for this task.
In a series of papers, the {possibility} was studied to {find}
some closed
analytical expressions
for one-loop $2-$, $3-$ and
$4-$ point functions for arbitrary dimension, external momenta
and masses (in principle also complex ones) in terms of generalized
hypergeometric functions with relatively simple integral
representations
\cite{Tarasov:2000sf,Fleischer:2002gs,hgf}.
The results immediately apply to a deduction of the coefficient $m_1$.
Such formulae} are also of great importance in the search of efficient
algorithms for the calculation of $5-$, $6-$ and higher point functions since
these functions may be reduced to $4-$ and lower point functions with
``unphysical'' external kinematics.

In this contribution, we explicitely perform
the  $\varepsilon$- expansion of the most complicated part:
The scalar one-loop box function $I_{1111}$ with two
photons (taken here in the s-channel), as it is needed for the
calculation of Bhabha
scattering up to order ${\cal O}(\ep)$.
Our starting point is the analytical expression as known from
\cite{Fleischer:2002gs,hgf}:
{
\begin{eqnarray}
&&\frac{(t-4m^2)}{  \Gamma \left(2-\frac{d}{2}\right) }
 I_{1111}^{(d)} = \nonumber \\
&&
{\frac{t-4m^2}{    i\pi^{d/2}\Gamma \left(2-\frac{d}{2}\right)}
\int \frac{d^dk_1}{k_1^2 (k_1^2+2q_4k_1)(k_1+q_1+q_4)^2
(k_1^2-2q_3k_1)}=}
\nonumber \\
&& - \frac{4 m^{d-4}}{s}
 F_2\left(\frac{d-3}{2},1,1,\frac32, \frac{d-2}{2};
 \frac{t}{t-4m^2}, -m^2Z \right)
 \nonumber \\
&&+\frac{4 m^{d-4}}{(d-3)s}
F^{1;2;1}_{1;1;0} \left[^{\frac{d-3}{2}:~ \frac{d-3}{2},~1;~~~~ 1;}
_{\frac{d-1}{2}:~~~~~~ \frac{d-2}{2};~~-;}~~-m^2Z,1-\frac{4m^2}{s}\right]
\nonumber \\
&& -\frac{\sqrt{\pi} (-s)^{\frac{d-4}{2}}}{2^{d-4}m  \sqrt{s}}
~\frac{\Gamma\left(\frac{d-2}{2}\right)}
{ \Gamma\left( \frac{d-1}{2}\right)}
F_1\left(\frac{d-3}{2},1,\frac12; \frac{d-1}{2};\frac{sZ}{4},1-\frac{s}{4m^2}
\right) 
\label{solution}
\end{eqnarray}
}
with ${Z=\frac{4 u}{s(4 m^2-t)},q_i^2=m^2,(q_1+q_4)^2=s,(q_1+q_2)^2=t
}$ and $s,t,u$ the usual Mandelstam variables.
Here {$F_1$, $F_2$ are Appell hypergeometric functions}
\begin{eqnarray}
F_1\left(\frac{d-3}{2},1,\frac{1}{2},\frac{d-1}{2};x,y\right)=
&&      \sum_{r=0}^{\infty} \sum_{s=0}^{\infty}
\frac{\left( \frac{d-3}{2} \right)_{r+s}}
{\left( \frac{d-1}{2} \right)_{r+s}}~\frac{\left( \frac{1}{2} \right)_s}
{\left( 1 \right)_s} ~x^r y^s, 
\end{eqnarray}
\begin{eqnarray}
F_2\left(\frac{d-3}{2},1,1,\frac{3}{2},\frac{d-2}{2};x,y\right)=
&&      \sum_{r=0}^{\infty} \sum_{s=0}^{\infty}
\frac{\left( \frac{d-3}{2} \right)_{r+s}}
{\left( \frac{3}{2} \right)_r \left( \frac{d-2}{2} \right)_s} ~x^r y^s 
\end{eqnarray}
and the Kamp\'e de F\'eriet function (KdF)  \cite{AppellKdF} is defined as
\begin{eqnarray}
F^{1;2;1}_{1;1;0} \left[^{\frac{d-3}{2}:~ \frac{d-3}{2},~1;~~~~ 1;}
_{\frac{d-1}{2}:~~~~~ \frac{d-2}{2};~~-;}~~x,y\right]=\!\!
&&
\sum_{r=0}^{\infty} \sum_{s=0}^{\infty}
\frac{\left( \frac{d-3}{2} \right)_{r+s}}
{\left( \frac{d-1}{2} \right)_{r+s}}~\frac{\left( \frac{d-3}{2} \right)_r}
{\left( \frac{d-2}{2} \right)_r} ~x^r y^s . 
\end{eqnarray}
The $\varepsilon-$ expansion of the generalized hypergeometric functions
occurring above is not quite straight forward. In particular since there
stands the factor
$ \Gamma \left(2-\frac{d}{2}\right) \sim \frac{1}{\ep}$ in front of all of them,
one needs their expansion up to order ${\ep}^2$. We have to develope
different techniques for each of them.

\section{Expansion of $F_1$}
We need to know the expansion
\begin{eqnarray}
F_1(\frac{d-3}{2},1,\frac{1}{2},\frac{d-1}{2};x,y)=F_1^0+\ep F_1^1+{\ep}^2 F_1^2+\cdots
\end{eqnarray}
with the kinematics: $x=-\frac{u}{t-4 m_e^2}<0, y=1-\frac{s}{4 m_e^2}<0, |y| \gg 1$~.
In two steps we obtain a form in which one of the {parameters} of $F_1 \sim \ep$.
The transformations are the following:
{
\begin{eqnarray}
&&F_1\left(\frac{d-3}{2},1,\frac{1}{2},\frac{d-1}{2};x,y\right)=
2 \frac{\Gamma(\frac{d-1}{2}) \Gamma(\frac{6-d}{2})}
{\Gamma(\frac{1}{2})(-y)^{\frac{d-3}{2}}}
{}_2F_1 \left[ 1, \frac{d-3}{2}, \frac{3}{2}; 1- z \right]
\non \\
&&~~~~~~~~~~~~~~~~+ ~\frac{d-3}{(d-6)(-x)\sqrt{-y}}
F_1\left(\frac{6-d}{2},1,\frac{1}{2},\frac{8-d}{2};\frac{1}{x},
\frac{1}{y}\right)
\end{eqnarray}
}
($\frac{x}{y} \equiv z = 1- \frac{s t}{(s-4 m_e^2)(t-4 m_e^2)};~~ 0 < z, 1-z < 1$)~~ and
{
\begin{eqnarray}
&&F_1\left(\frac{6-d}{2},1,\frac{1}{2},\frac{8-d}{2};\frac{1}{x},
\frac{1}{y}\right)
= \frac{\Gamma(\frac{1}{2})~\Gamma(\frac{8-d}{2})}
{\Gamma(\frac{7-d}{2})}y^{\frac{6-d}{2}}
~{}_2F_1 \left[ 1, \frac{6-d}{2}, \frac{7-d}{2}; \frac{1}{z}\right]  \non \\
&&~~~~~~~~~~~~~~~
+\frac{(d-6)\sqrt{X-1}~(Y-X)}{\sqrt{X}} F_1\left(1,\frac{d-4}{2},1,
\frac{3}{2};X,Y\right)
\end{eqnarray}
}
with $X=1-y=\frac{s}{4 m_e^2} \gg 1, Y=\frac{y-1}{x-1}=1-\frac{t}{4 m_e^2} \gg 1~~
(X > Y,~ 1-\frac{1}{Y}=\omega)$. Here we observe that the argument of the $~_2F_1$
as well as those of the $F_1$ function are larger than 1, i.e. both functions are complex
and the imaginary parts must cancel since the $F_1$ on the l.h.s. has arguments less
than 0 and thus is real. For the imaginary part of the $F_1$ function we obtain
{
\begin{eqnarray}
&&{\rm Im}~F_1\left(1,\frac{d-4}{2},1,\frac{3}{2};X,Y\right)
=\non \\
&&~~~~~~~~~~~~~~~\frac{{\pi}^{\frac{3}{2}}\sqrt{X}(X-1)^{\frac{3-d}{2}}}
{2 Y \Gamma(\frac{5-d}{2})
\Gamma(\frac{d-2}{2})} ~_2F_1 \left[ 1, \frac{d-3}{2}, \frac{d-2}{2}; z\right].
\end{eqnarray}
}
Transforming $~_2F_1 \left[ 1, \frac{6-d}{2}, \frac{7-d}{2}; \frac{1}{z}\right]$ with
argument $\frac{1}{z} > 1$ to a $~_2F_1$ function with argument $z < 1$, one shows
that the imaginary parts cancel. Transforming (the real part) further to the
argument $1-z$ one finally obtains
\begin{eqnarray}
&&F_1\left(\frac{d-3}{2},1,\frac{1}{2},\frac{d-1}{2};x,y\right)=
-(d-3)\frac{Y}{\sqrt{X}}~{\rm Re}~ F_1\left(1,\frac{d-4}{2},1,\frac{3}{2};X,Y\right) \non \\
&& +(d-3)\frac {\Gamma(\frac{d-3}{2}) \Gamma(\frac{6-d}{2})}
{\Gamma(\frac{1}{2})(-y)^{\frac{d-3}{2}}} {\rm sin}^2(\pi \frac{d}{2})
~_2F_1 \left[ 1, \frac{d-3}{2}, \frac{3}{2}; 1- z \right] \non \\
&&-(d-3)\frac{\pi}{2} {\rm sin}(\pi \frac{d}{2}) (-x)^{-\frac{d-4}{2}} \frac{1}{\sqrt{-y(1-z)}}.
\end{eqnarray}
To expand up to the required order ($ \sim {\varepsilon}^2$),
it is sufficient to set
${\rm sin}(\pi \frac{d}{2})=-\pi \varepsilon$ and
\begin{eqnarray}
~_2F_1 \left[ 1, \frac{d-3}{2}, \frac{3}{2}; 1- z \right]=\frac{1}{2
  \sqrt{1-z}}
\Lg(\frac{1+\sqrt{1-z}}{1-\sqrt{1-z}}) {+O(\varepsilon)}  .
\end{eqnarray}
This simplifies the expansion
considerably and the ${\rm Re}~ F_1\left(1,\frac{d-4}{2},1,\frac{3}{2};X,Y\right)$
we take from \cite{hgf}. Characteristic variables appearing in the result
are
\begin{eqnarray}
   A=\frac{\sqrt{1-\frac{1}{X}}-1}{\sqrt{1-\frac{1}{X}}+1} ~< ~0,
~~~B=\frac{\sqrt{1-\frac{1}{Y}}-1}{\sqrt{1-\frac{1}{Y}}+1} ~< ~0.
\end{eqnarray}
and introducing $a=\sqrt{1-\frac{1}{X}}$,
$b=\sqrt{1-\frac{1}{Y}}$ we can write ($1> a > b > 0$ )
$A=\frac{a-1}{a+1}, B=\frac{b-1}{b+1}$ and
\begin{eqnarray}
F_1^0 =-\frac{m_e}{\sqrt{s}}\frac{1}{b} \Lg(-B),
\end{eqnarray}
yielding the correct $\frac{1}{\ep}$-term of $D_0$ {\cite{tHooftVeltman}}.
Keeping only the leading terms, collecting the contributions,
yields correspondingly
{
\begin{eqnarray}
&&\frac{2}{s (t - 4m_e^2)}\frac{\Gamma(\frac{1}{2})}
{\Gamma(\frac{1}{2}-\ep)}
\Gamma(1-\ep) \left(-\frac{s}{4}\right)^{-\ep}
\Gamma(\ep)\frac{1}{b}
 \Bigl[ {\rm Re} \left\{ \Lg(B) + \cdots  \right\} \Bigr.
 \non \\
&&~~\left. -{\pi}^2 {\ep}^2 \Lg(\frac{1- A B}{A-B})
               -{\pi}^2 \ep (1+ \ep \Lg(\frac{X}{Y}-1))
+2\ep^2\pi^2
	       +O({\ep^3}) \right],
\end{eqnarray}
}
where the higher order terms in $\ep$ of ${\rm Re} \left\{ \Lg(B) + \cdots  \right\}$
have to be taken from \cite{hgf}.

\section{Expansion of $F_2$}
We need to know the expansion
\begin{eqnarray}
F_2(\frac{d-3}{2},1,1,\frac{3}{2},\frac{d-2}{2};\omega,z)=F_2^0+\ep F_2^1+{\ep}^2 F_2^2+\cdots
\end{eqnarray}
with the kinematics: $\omega=\frac{t}{t-4 m_e^2},
z=-4 m_e^2\left(\frac{1}{s}+\frac{1}{t-4 m_e^2}\right)$. At first we perform
the following Euler transformation:
{
\begin{eqnarray}
&&F_2 \Bigl(\frac{d-3}{2},1,1,\frac{3}{2},\frac{d-2}{2};\omega,z
\Bigr)=
\non \\
&&~~~~~~~~~
(1-z)^{\frac{3-d}{2}} F_2 \Bigl(\frac{d-3}{2},1,\frac{d-4}{2},
\frac{3}{2},\frac{d-2}{2};
\frac{\omega}{1-z},-\frac{z}{1-z}\Bigr).
\end{eqnarray}
}
The factor $(1-z)^{-\frac{d-3}{2}}$ will be dropped in
what follows and is taken into account again when collecting the results.
Introducing $\frac{\omega}{1-z}={\omega}^{'}$ and $-\frac{z}{1-z}={z}^{'}$, we have
\begin{eqnarray}
0 \le {\omega}^{'}+{z}^{'} < 1.
\end{eqnarray}
With
$\alpha=\frac{1}{2}-\varepsilon, \beta=1, {\beta}^{'}=-\varepsilon, \gamma=\frac{3}{2}$
and ${\gamma}^{'}=1-\varepsilon$
\begin{eqnarray}
F_2(\alpha,\beta,{\beta}^{'},\gamma,{\gamma}^{'},{\omega}^{'},{z}^{'})
=_2F_1(\alpha,\beta,\gamma;{\omega}^{'})
+ {\beta}^{'} S(\alpha,\beta,{\beta}^{'},\gamma,{\omega}^{'},{z}^{'})~, 
\end{eqnarray}
where ${\gamma}^{'}=1+{\beta}^{'}$ has been used and
\begin{eqnarray}
S(\alpha,\beta,{\beta}^{'},\gamma,{\omega}^{'},y)=\sum_{n=1}^{\infty}
\frac{(\alpha)_n}{{\beta}^{'}+n} \frac{{y}^n}{{n!}}
{~}_2F_1(\alpha+n,\beta,\gamma;{\omega}^{'})
\end{eqnarray}
In order to get rid of the denominator ${\beta}^{'}+n$ we differentiate S w.r.t. y
and use \cite{Prudnikov:1986}
\begin{eqnarray}
\sum_{n=0}^{\infty} \frac{(\alpha)_n (\delta)_{n}} {{n!}{(\delta}^{'})_n} y^n
{~}_2F_1(\alpha+n,\beta,\gamma;{\omega}^{'})=
F_2(\alpha,\beta,\delta,\gamma,{\delta}^{'};{\omega}^{'},y)
\end{eqnarray}
with $\delta={\delta}^{'}$. Applying again the Euler relation,
\begin{eqnarray}
F_2(\alpha,\beta,\delta,\gamma,\delta;{\omega}^{'},y)
&& =(1-y)^{-\alpha} F_2(\alpha,\beta,0,\gamma,\delta;\frac{{\omega}^{'}}{1-y},-\frac{y}{1-y})
\nonumber \\
&& =(1-y)^{-\alpha} {~}_2F_1(\alpha,\beta,\gamma;\frac{{\omega}^{'}}{1-y}),
\end{eqnarray}
we finally have ($\beta=1, \gamma=\frac{3}{2}$ inserted)
\begin{eqnarray}
S(\alpha,{\beta}^{'},{\omega}^{'},{z}^{'})
&& =
\int_{y=0}^{{z}^{'}}\frac{\partial S(\alpha,{\beta}^{'},{\omega}^{'},y)}
{\partial y} dy \non \\
&& =S_0(\frac{1}{2},0,{\omega}^{'},{z}^{'})
+\ep S_1(\frac{1}{2},0,{\omega}^{'},{z}^{'})
{+O(\varepsilon^2)},
\end{eqnarray}
with
{
\begin{eqnarray}
&&\frac{\partial S}{\partial y}=
\frac{1}{y}\left[\frac{1}{(1-y)^{\alpha}}
{~}_2F_1(\alpha,1,\frac{3}{2};\frac{{\omega}^{'}}{1-y})
-{~}_2F_1(\alpha,1,\frac{3}{2};{\omega}^{'})\right]
\non \\
&&~~~~~~~~~~~~~~~~~+
\frac{\varepsilon}{y} S(\frac{1}{2},0,{\omega}^{'},y)+O({\varepsilon}^2).
\end{eqnarray}
}
To complete the $\varepsilon$-expansion, we need (with $z=\frac{{\omega}'}{1-y}$)
{
\begin{eqnarray}
&&{~}_2F_1(\alpha,1,\frac{3}{2};z)={~}_2F_1(\frac{1}{2},1,\frac{3}{2};z)
\non \\
&&~~~~~~~~~~~~~+
\varepsilon {\delta}^{(1)}F(z,u)+{\varepsilon}^2 {\delta}^{(2)}F(z,u)+\cdots
\end{eqnarray}
}
with the abbreviation $u=\frac{1+\sqrt z}{1-\sqrt z}$. Explicitely
$u=\frac{\sqrt{1-y}+w}{\sqrt{1-y}-w}$ with $w=\sqrt{{\omega}'}$,
\begin{eqnarray}
{\delta}^{(1)}F(z,u)=\frac{1}{2 \sqrt z}
[2 {\rm Li_2}(-\frac{1}{u})-2 \Lg(u) \Lg(1+u)+\frac{3}{2}{\Lg}^2(u) +\zeta(2) ]
\end{eqnarray}
and
{
\begin{eqnarray}
&&{\delta}^{(2)}F(z,u)
=\frac{1}{2 \sqrt z}
\left[-4 S_{1,2}(-\frac{1}{u})-(\Lg(u)+2 \Lg(1+\frac{1}{u})) \zeta(2)
\right.
\non \\
&&~~~-4 \Lg(1+\frac{1}{u}) {\rm Li_2}(-\frac{1}{u})
+2 \Lg(u) {\Lg}^2(1+\frac{1}{u})
+{\Lg}^2(u) \Lg(1+\frac{1}{u}) \non \\
&&~~~\left. +\frac{1}{6}{\Lg}^3(u)+2 \zeta(3)+
2 {\rm Li_3}(-\frac{1}{u}) \right].
\end{eqnarray}
}
\subsection{{Order $\ep$ of $F_2$}}
In this order we have
\begin{eqnarray}
F_2(\alpha,\beta,{\beta}^{'},\gamma,{\gamma}^{'},{\omega}^{'},{z}^{'})
&&={~}_2F_1(\frac{1}{2},1,\frac{3}{2};{\omega}^{'})
\nonumber \\
&&+\ep {\delta}^{(1)}F({\omega}^{'},u_0)
-\ep S(\frac{1}{2},0,{\omega}^{'},{z}^{'})  +
O(\varepsilon^2),
\nonumber \\
&& \equiv F_2^0+\ep F_2^1 {+ O(\varepsilon^2)},
\end{eqnarray}
where the `scale' $ u_0=u(y=0)
=\frac{1+w}{1-w} \sim \frac{s}{m_e^2} \frac{1}{1-\frac{4 m_e^2}{t}}. $
$u_0$ is large for $s \gg  m_e^2$ and $-t \gg 4  m_e^2$ and sets the scale for the variable
u in general. Further we introduce  $u_1=u(y={z}^{'}) < u_0$. Thus we can write
\begin{eqnarray}
F_2^0
&&=\frac{1}{2w} \Lg(u_0), \non \\
S(\frac{1}{2},0,{\omega}^{'},y)
&& =\frac{1}{2w}\int_{y=0}^{y}\frac{1}{y}\Lg(\frac{u}{u_0}) dy \nonumber \\
&& =\frac{1}{2w}\int_{u}^{u_0}\left[\frac{2}{u-1}-\frac{1}{u-u_0}-
\frac{1}{u-\frac{1}{u_0}}\right]\Lg(\frac{u}{u_0}) du \nonumber \\
&& \equiv \frac{1}{2 w} S_0(u_0,u).
\end{eqnarray}
and $S(\frac{1}{2},0,{\omega}^{'},{z}^{'})=\frac{1}{2w}S_0(u_0,u_1)$.
The integration yields:
{
\begin{eqnarray}
&&S_0(u_0,u)=
 2{\rm Li_2}(\frac{1}{u})+{\rm Li_2}(\frac{u}{u_0})-{\rm Li_2}(\frac{1}{u_0 u})
+2{\rm Li_2}(-\frac{1}{u_0})-\zeta(2)- \nonumber \\
&& ~~~~\Lg(\frac{u}{u_0})\left[2\Lg(u-1)-\Lg(u_0 u-1)
-\Lg(1-\frac{u}{u_0})-\frac{1}{2}\Lg(\frac{u}{u_0} {)} \right].
\end{eqnarray}
}
\subsection{{Order ${\ep}^2$ of $F_2$}}
The next order can be written in the form
\begin{eqnarray}
F_2^2={\delta}^{(2)}F({\omega}^{'},u_0)-S_1(\frac{1}{2},0,{\omega}^{'},{z}^{'})
\end{eqnarray}
with
{
\begin{eqnarray}
&&S_1(\frac{1}{2},0,{\omega}^{'},{z}^{'})
\equiv S_1(u_0,u_1)=
\int_{y=0}^{{z}^{'}} \frac{dy}{y}
\left\{\frac{1}{2 w} \Lg (1-y)~\Lg (u)
\right.
\non \\
&&~~~~~~\left. +{\delta}^{(1)}F({\omega}^{'},u)
-{\delta}^{(1)}F({\omega}^{'},u_0)+
S(\frac{1}{2},0,{\omega}^{'},y)\right\}
\end{eqnarray}
}
The curly bracket in the above integrand finally reads
{
\begin{eqnarray}
&&\left\{ \cdots \right\}=
 {\rm Li_2}(\frac{u}{u_0})-{\rm Li_2}(1)+
{\rm Li_2}(\frac{1}{u^2})-{\rm Li_2}(\frac{1}{u_0 u})+
2\Lg(u)\Lg(\frac{u_0-1}{u-1}) \non \\
&&~~~~~~~
-2\Lg(u_0+1) \Lg(\frac{u}{u_0})
+ \frac{3}{2}\Lg(u_0 u) \Lg(\frac{u}{u_0})-\Lg(\frac{u}{u_0})
\non \\
&&~~~~~~~
\left[2\Lg(u-1)-\Lg(u_0 u-1)-\Lg(1-\frac{u}{u_0})
-\frac{1}{2}\Lg(\frac{u}{u_0})\right]
\end{eqnarray}
}
There is no problem to perform the final integration, but the expressions
blow up considerably. Therefore we confine ourselves here to the leading terms
only by considering $u $ $(u_0)$ as large and drop the small quantities.
Then the integral can be written in the simplified form ($\frac{u}{u_0}=v$ the
new integration variable)
{
\begin{eqnarray}
&&S_1(u_0,u_1)
=\frac{1}{2 w}\int_{v=r}^{1}\left[\frac{1}{v}+\frac{1}{1-v}\right]
\Bigl\{{\rm Li_2}(v)-{\rm Li_2}(1)-{\Lg}^2(v)-\Lg(u_0)\Lg(v)
\Bigr. \non \\
&&\Bigl.
+\Lg(v)\Lg(1-v)\Bigr\}
=\frac{1}{2 w} \Bigl[-{\rm Li_3}(1-r)+\Lg(1-r){\rm Li_2}(r)
\Bigr. \non \\
&&
+(\Lg(u_0)+\Lg(1-r)){\rm Li_2}(1-r)
-\Lg(r){\rm Li_2}(r)+\frac{1}{3}{\Lg}^3(r)
\non \\
&&\left.
+(\frac{1}{2}\Lg(u_0)-\Lg(1-r)){\Lg}^2(r)
+{\Lg}^2(1-r) \Lg(r)-\zeta(2)\Lg(\frac{1}{r}-1) \right]
\end{eqnarray}
}
with $0 < r=\frac{u_1}{u_0} <1$. If one wants higher precision, it is easier to
expand in $\frac{1}{u} (\frac{1}{u_0})$ instead of performing all integrals
analytically, which is possible nevertheless.
Collecting the results, we have
\begin{eqnarray}
-\frac{2 (m_e^2)^{-\ep}}{s(t-4 m_e^2)}(1-z)^{\ep}
\Gamma(\ep)\frac{1}{\sqrt{\omega}}
\left(\Lg(u_0)+\cdots \right)
\end{eqnarray}
\section{Expansion of the  Kamp\'e de F\'eriet function}
We need to know the expansion
\begin{eqnarray}
F^{1;2;1}_{1;1;0} \left[^{\frac{d-3}{2}:~ \frac{d-3}{2},~1;~~~~ 1;}
_{\frac{d-1}{2}:~~~~~~ \frac{d-2}{2};~~-;}~~x,y\right]
=K^0+\ep K^1+{\ep}^2 K^2+\cdots. 
\end{eqnarray}
with the kinematics: $x=-4 m_e^2\left(\frac{1}{s}+\frac{1}{t-4 m_e^2}\right),
y=1-\frac{4 m_e^2}{s}$.
In this case we begin with the integral representation of the KdF function:
{
\begin{eqnarray}
&&F^{1;2;1}_{1;1;0} \left[^{\frac{d-3}{2}:~ \frac{d-3}{2},~1;~~~~ 1;}
_{\frac{d-1}{2}:~~~~~~ \frac{d-2}{2};~~-;}~~x,y\right]
\non \\
&&~~~~~~~~=\frac{d-3}{2}\int_0^1\frac{dt~ t^{\frac{d-5}{2}}}{1-t~ y}
{~}_2F_1(1,\frac{d-3}{2},\frac{d-2}{2},x~ t)
\end{eqnarray}
}
Again we perform a shift such that one of the parameters of the ${~}_2F_1 \sim \ep$:
\begin{eqnarray}
{~}_2F_1(1,\frac{d-3}{2},\frac{d-2}{2},x~ t)
&& =(1-x~ t)^{-\frac{1}{2}}{~}_2F_1(-\ep,\frac{1}{2},1-\ep,x~ t) \non \\
&& =(1-x~ t)^{-\frac{1}{2}}\left[1 - \ep S(x~t)\right].
\end{eqnarray}
with
\begin{eqnarray}
S(x~t)=\sum_{n=1}^{\infty}\frac{1}{n-\ep}\frac{(\frac{1}{2})_n}{n!}(x~t)^n.
\end{eqnarray}
The $\ep$-expansion of $S(x~t)$ can again be obtained by first
differentiating $S$:
\begin{eqnarray}
\frac{\partial S}{\partial x}
&&=\frac{1}{x}\left(\frac{1}{\sqrt {1-x~t}}-1\right)
+\frac{\ep}{x}\sum_{n=1}^{\infty}\frac{1}{n}\frac{(\frac{1}{2})_n}{n!}(x~t)^n +O({\ep}^2)\non \\
&&=\frac{1}{x}\left(\frac{1}{\sqrt {1-x~t}}-1\right)+\frac{\ep}{x} S(x~t)|_{\ep=0}+O({\ep}^2).
\end{eqnarray}
In order to obtain S to $O(\ep)$, we need the following integrals:
\begin{eqnarray}
S(x~t)|_{\ep=0}=\int_0^x\frac{dx}{x}\left(\frac{1}{\sqrt {1-x~t}}-1\right)=2 \Lg(1+\frac{1}{v})
\end{eqnarray}
and
\begin{eqnarray}
\int_0^x\frac{dx}{x}S(x~t)|_{\ep=0}=-2 {\rm Li_2}(-\frac{1}{v})-2{\Lg}^2(1+\frac{1}{v}),
\end{eqnarray}
where we introduced $v=\frac{1+\sqrt{1-x~t}}{1-\sqrt{1-x~t}}$.
Thus the above integral reads
\begin{eqnarray}
\int_0^1 \frac{dt~ t^{-\frac{1}{2}}}{1-t~ y}\frac{1}{\sqrt{1-x~t}}
\left\{1-\ep \Lg(\frac{4}{v~x})-
{\ep}^2 \left[-2 {\rm Li_2}(-\frac{1}{v})-\frac{1}{2}{\Lg}^2(\frac{4}{v~x})\right]
\right\}
\end{eqnarray}
After a variable transformation we can write
{
\begin{eqnarray}
&&\int_0^1 \frac{dt~ t^{-\frac{1}{2}}}{1-t~ y}\frac{1}{\sqrt{1-x~t}} f(v)
=-\frac{1}{\sqrt{y-x}}\int_0^1 dt \left\{
 b_1\left[\frac{1}{1+b_1~t}+\frac{1}{1-b_1~t}\right] \right. \non \\
&&~~~~~~~~~~~~
\left.
-b_2\left[\frac{1}{1+b_2~t}+\frac{1}{1-b_2~t}\right] \right\}f(\frac{v_1}{t^2})
\end{eqnarray}
}
with $v_1=v(t=1)=\frac{1+\sqrt{1-x}}{1-\sqrt{1-x}},~ b_1=\frac{1}{\sqrt{v_0~v_1}}$ and
$b_2=\sqrt{\frac{v_0}{v_1}}$ with $v_0=\frac{1+\sqrt{1-\frac{x}{y}}}{1-\sqrt{1-\frac{x}{y}}}$.
$v_0$ and $v_1$ both being large, results in $b_1 \ll 1$ and $b_2<1$ but very close to 1.
Taking again the attitude to keep only leading contributions, the $b_1$-contribution can be
dropped.
The ${\rm Li_2}$-function
in the second order term can be written as
\begin{eqnarray}
{\rm Li_2}(-\frac{t^2}{v_1})=2\left[{\rm Li_2}( i\frac{t}{\sqrt{v_1}})+
                                    {\rm Li_2}(-i\frac{t}{\sqrt{v_1}})\right]
\end{eqnarray}
so that integration is posible. We do get,however, relatively complicated
complex conjugate contributions. On the other hand since $v_1 \gg 1$ this contribution
is small from the very beginning and can be well approximated by expanding the
${\rm Li_2}$-function. Here it is dropped alltogether. Thus we are left with the
following contributions:
\begin{eqnarray}
K^0=\frac{d-3}{2 \sqrt{\omega} }~\Lg(u_0),
\end{eqnarray}
\begin{eqnarray}
K^1=\frac{d-3}{2 \sqrt{\omega} }\left[\Lg(\frac{1+b_2}{1-b_2})\Lg(\frac{v_1~x}{4})
+2({\rm Li_2}(b_2)-{\rm Li_2}(-b_2))\right]
\end{eqnarray}
{
\begin{eqnarray}
&&
K^2=\frac{d-3}{2 \sqrt{\omega} }
\left[\frac{1}{2} \Lg(\frac{1+b_2}{1-b_2}){\Lg}^2(\frac{v_1~x}{4})+
2 \Lg(\frac{v_1~x}{4})({\rm Li_2}(b_2)-{\rm Li_2}(-b_2)) \right. \non \\
&&
~~~~~~~~~~~~~~~~~~~~\Bigl. +4({\rm Li_3}(b_2)-{\rm Li_3}(-b_2))
\Bigr]
\end{eqnarray}
}
Collecting the results we obtain
\begin{eqnarray}
\frac{2 (m_e^2)^{-\ep}}{s(t-4 m_e^2)}\Gamma(\ep)\frac{1}{\sqrt{\omega}}
\left(\Lg(u_0)+\cdots \right)
\end{eqnarray}
We see that the $\frac{1}{\ep}$-term cancels against the one from the $F_2$ contribution.

{
\section{Expansion of the scalar box function with Feynman parameters}}
\newcommand{\ba}{\begin{eqnarray}}
\newcommand{\ea}{\end{eqnarray}}
\newcommand{\nl}{\nonumber \\}
In order to have a completely independent numerical check of the above
results, we derived a Feynman parameter integral representation for
the $\varepsilon$-expansion.
We follow closely
\cite{Berends:1973fd}, where the scalar four-point integral was treated
with a finite photon mass in $d=4$ dimensions.

The function to be calculated is, in LoopTools notations \cite{Hahn:1998yk}:
\ba
J &=& D_0(m^2,m^2,m^2,m^2~|~t,s~|~m^2,0,m^2,0)
\nl
&=& \frac{(2\pi\mu)^{2\varepsilon}}{i\pi^2}
~\int\frac{d^dk}{k^2(k^2+2kq_4)(k^2-2kq_3)(k+q_1+q_4)^2} .
\ea
A constant transforms the normalization of $D_0$ to that of
$I_{1111}^{(d)}$:
\ba
D_0 &=& (4\pi\mu^2)^{\varepsilon}~ I_{1111}^{(d)}.
\ea
The infrared singularity may be isolated in a 3-point function $C_0$ by
redefining
\ba\label{eq-4-1}
J &=& \frac{2}{s}~(F+C_0),
\ea
with
\ba
C_0 &=&  C_0(t,\mu^2,m^2~|~m^2,\mu^2,0) ~=~
\frac{(2\pi\mu)^{2\varepsilon}}{i\pi^2}
\int\frac{d^dk}{k^2(k^2+2kq_4)(k^2-2kq_3)} ,
\nl
\ea
and with
\ba
F&=& \frac{(2\pi\mu)^{2\varepsilon}}{i\pi^2}
\int\frac{d^dk~(s/4-k^2)}{k^2(k^2+2kq_4)(k^2-2kq_3)(k+q_1+q_4)^2}
\ea
being a finite scalar four point function.

The $\varepsilon$-expansions may be easily derived now
starting from
\ba
C_0 &=&
\Gamma(\varepsilon)\int_0^1 \frac{dx}{2p_{x}^2}
\left[\frac{4\pi\mu^2}{p_{x}^2}\right]^{\varepsilon} ,
\label{eq-1248}
\\
F &=&\Gamma(2+\varepsilon) \int_0^1 dx dy dz
\frac{y^2z}{(M^2)^2}
~ \left(\frac{1}{2}yzs + \frac{1-2\varepsilon}{1+\varepsilon}M^2\right)
\left[\frac{4\pi\mu^2}{p_{x}^2}\right]^{\varepsilon}
\nl
&=& \Gamma(2+\varepsilon) \left[ I_0 + \varepsilon I_1 +
  \varepsilon I_L \right] + \ldots,
\ea
with
\ba
p_x^2 &=& -x(1-x)t + m^2 -i\epsilon,
\\
M^2 &=& y[y z^2 p_x^2 - (1-y)(1-z)s].
\ea
Thus, the four-point function may be represented as follows: 
\ba\label{eq-irdiv3-a}
I_0 &=&
- \int_0^1 \frac{dx}{2p_x^2} \ln\left(-A\right), 
\\
I_1 &=&
{-3}\int_0^1 \frac{dx}{p_x^2} dz
\left[
\frac{z}{N(z)}
+ \frac{Az(1-z)}{N(z)^2} \ln\frac{z^2}{(1-z)(-A)}
\right] ,
\ea
with
\ba
A&=&\frac{s}{p_x^2},
\\
N(z)&=&z^2+(1-z)A.
\ea
The last function will be given here in short as a three-fold
integral.
 But it is evident that the $y$-integration leads to simple
integrals in terms of dilogarithms or simpler functions:
\ba
I_L &=& \int_0^1 \frac{dx}{p_x^2} dz dy
\left[
\frac{Ay}{2z^2K(y)^2}
+ \frac{y}{zK(y)} \right]\ln\frac{4\pi\mu^2A}{z^2yK(y)s} ,
\ea
with
\ba
K(y)&=&y-(1-y)\frac{1-z}{z^2}A.
\ea

\end{document}